# A Heliosheath Model for the Origin of the CMB Quadrupole Moment

H.N. Sharpe

Bognor, Ontario, Canada
sh3149@brucetelecom.com

## ABSTRACT

A non-cosmological origin for the CMB quadrupole moment is suggested in this paper. Geometric distortions to an otherwise isotropic CMB could be imprinted on the CMB radiation as it propagates through the asymmetric termination shock formed at the boundary of the solar wind and the local interstellar medium. In addition to this boundary distortion, the Voyager spacecraft observed abrupt changes in plasma properties and rapidly fluctuating magnetic and electric fields as they recently crossed the termination shock and entered the heliosheath. Several mechanisms are discussed which could potentially imprint the termination shock distortion on the CMB. Temporal variations of this distortion due to solar wind pressure changes could manifest in the multipole moments of the CMB. Speculations are presented for the effect of heliosheath radiative and dynamical processes on the observed small-scale angular power spectrum of the CMB.

## INTRODUCTION

The cosmic microwave background radiation (CMB) is predominantly isotropic. Large scale and small scale anisotropies observed in the angular power spectrum are usually given a cosmic interpretation [1]. The two lowest harmonics, the dipole and quadrupole moments are the most interesting large-scale anisotropies in the CMB for cosmology. It has been established that the CMB dipole is caused by the motion of the observer relative to the CMB rest frame. But the quadrupole (and higher order) moment must be produced by a different mechanism. The origin of the CMB quadrupole has been the subject of intense cosmological research [2]. Differential redshifts caused by anisotropic expansion is one important example.

It is the purpose of the present note to suggest that the CMB quadrupole may in fact have a more local origin. Although originally believed to be perfectly spherical, recent results from the two Voyager spacecraft (V1,V2) indicate that the solar wind termination shock (TS) which surrounds our solar system is asymmetric [3]. The TS forms where the outflowing supersonic solar wind is slowed to subsonic through its interaction with the interstellar wind. The TS marks the inner boundary of the heliosheath. The heliosheath is characterized by a turbulent, magnetized plasma [4]. The outer boundary is the heliopause and occurs where the solar wind pressure balances that of the interstellar medium.

Analysis of the V1, V2 data collected while these spacecraft crossed the TS into the heliosheath shows the TS to be a boundary of abrupt changes in pressure, temperature, density, magnetic and electric field properties and ion and electron (plasma) properties [5,6]. It should therefore be expected that the TS (and heliosheath) will leave an imprint on observed radiation which originates outside of the solar system and propagates through this optically thin boundary. We suggest that this imprint is in fact the observed quadrupole moment in the CMB. For this analysis the CMB is assumed to be perfectly isotropic outside the heliosphere. In a sense the TS introduces an "astigmatism" into the CMB.

In the next section we derive an expression for the quadrupole in an idealized TS and relate it to the observed CMB value. This is followed by a discussion of mechanisms and implications for the angular power spectrum of the small-scale CMB fluctuations.

## MODEL

We represent the TS by an idealized prolate ellipsoid of revolution; a:a:c where c > a. The equation for the surface is:

$$r(\theta,\phi) = a^2 c[\, a^2 c^2 \sin^2\theta + a^4 \cos^2\theta\, ]^{-1/2} \tag{1}$$

The usual spherical harmonic expansion expresses (1) as:

$$r(\theta,\phi) = \sum_{n=0}^{\infty} \sum_{m=-n}^{n} C_{nm}\, Y_n^m(\theta,\phi) \tag{2}$$

The rms amplitude for the nth moment is:

$$r_n = \left[ \frac{1}{4\pi} \sum_{m=-n}^{n} C_{nm}^2 \right]^{1/2} \tag{3}$$

From symmetry only those coefficients with m=0 are nonzero. Also, because (1) is even in $\theta$, there is no dipole moment. We then derive the following expression for the ratio of the quadrupole moment to the monopole moment for this surface:

$$\frac{r_2}{r_0} = \frac{C_{20}}{C_{00}} = \frac{3\sqrt{5}}{4}\left[ \frac{1}{f^2} - 2\left( \frac{1}{3} + \frac{1}{g\sqrt{\left(\frac{c}{a}\right)^2 - 1}} \right) \right] \tag{4}$$

where: $f \equiv \sqrt{1 - \left(\frac{a}{c}\right)^2}$  and  $g \equiv \pi - 2\cos^{-1} f$

The CMB temperature anisotropies may also be represented by (3), where $r_n$ is replaced by $T_n$. The observed constraint on the ratio of the temperature quadrupole to monopole moments is [1,12]:

$$\frac{T_2}{T_0} \approx 5x10^{-6} \qquad (5)$$

In the CMB spherical harmonic analysis, all the $C_{nm}$ in (3) are generally nonzero. However in the present simple model we wish to determine the ratio (a/c) in (4) which accounts for this observed CMB quadrupole. Equating (4) and (5) then we find that (a/c) = 0.95.

Richardson et al [5] report a N-S asymmetry for the TS of 7-8 AU with a variability of 2-3 AU due to solar wind pressure changes. This distortion is comparable to that just computed for the prolate ellipsoid assuming an effective mean radius of 90 AU for the TS. Clearly these results are not conclusive as the TS is not an idealized ellipsoidal surface with zero dipole moment. They do suggest however the possibility that the CMB quadrupole moment may be imprinted by the geometric distortion of the TS. Next we consider possible mechanisms for coupling of the CMB to this TS region.

## MECHANISMS

There are numerous physical processes occurring at the TS (and within an optically thin heliosheath) which could modify an otherwise isotropic background CMB. They fall into two broad categories: those mechanisms which directly interact with the CMB radiation and those which generate radiation within the optically thin TS region which adds to the observed CMB blackbody spectrum (possibly modifying it).

Thomson scattering of the CMB due to enhanced electron density at the TS belongs to the first category. The heliosheath is optically thin to this radiation. Compton and possibly inverse-Compton scattering are other possible mechanisms in the energetic plasma environment near the TS. Intense localized plasma-wave electric fields recently observed at the TS [6] could accelerate electrons to relativistic energies. If the TS plasma is dusty due to the entrainment of solar system particles by the solar wind then Rayleigh scattering may also occur. Another possibility is a modification to the index of refraction near the TS. Since ($I_\nu/n_r^2$) is constant along a CMB ray through the TS [10], ($I_\nu$ is the specific intensity and $n_r$ is the index of refraction) the CMB would suffer a distortion which mimics the shape of the TS, thereby acquiring a quadrupole moment. However at WMAP frequencies this effect is negligible though the observed sharp increases in the temperature, plasma density and magnetic field strength across the TS [5] may produce an additional effect on the refractive index. We are currently studying this problem.

Synchrotron radiation due to electron acceleration along heliosheath magnetic field lines [8] and thermal bremsstrahlung emission at the TS are two in situ processes which could emit radiation that conforms to the geometric distortion of the TS. From the V1, V2 data the optical depths for these mechanisms is small in the heliosheath. The radiation then will just add to the unattenuated CMB background. This problem is currently being studied.

Some of these mechanisms have the potential to modify the CMB black body spectrum. These include Compton and Rayleigh scattering and synchrotron emission. The latter may also introduce polarization effects. Recent research has shown the CMB spectrum to possess small low order harmonic non-black body components [7,13]. We suggest that these moments could originate from the TS distortion. At WMAP frequencies and the TS electron temperatures, thermal bremsstrahlung emission is essentially independent of frequency.

All these mechanisms may also have relevance to the observed power spectrum of the small-scale anisotropy of the CMB. This power spectrum is usually interpreted in terms of primordial fluctuations at the surface of last scattering and has lead to the standard flat $\Lambda CDM$ cosmology. But if in fact the heliosheath mechanisms listed above are relevant, then we may be viewing the cosmos through a "dirty" optically thin heliosheath "window". The heliosheath plasma is modeled with MHD codes [9, 11] which have the additional possibility of developing magnetoacoustic waves. These acoustic waves could mimic the acoustic ripples seen in the CMB analysis. Polarization of the heliosheath radiation is also expected due to synchrotron processes. This discussion is speculative. On-going research is focused on determining the impact that heliosheath radiative and dynamical processes may have on the observed angular power spectrum of the CMB small-scale anisotropies.

Finally we mention that the TS is expected to change shape with the solar wind pressure [5]. Any observed temporal changes of the CMB power spectrum would then be a good test of the model proposed in this paper. Recently temporal low harmonic residuals have been reported from a comparison of the WMAP3 and WMAP5 maps [12].